\documentclass[aps,prd,twocolumn,showpacs,superscriptaddress,groupedaddress]{revtex4}
\usepackage{graphicx}
\usepackage{dcolumn} 
\usepackage{bm}      
\usepackage{amssymb}

\newcommand{\mbb}{$m_{\beta\beta}$}
\newcommand{\iso}[2]{\ensuremath{^{#2}\textrm{#1}}}
\newcommand{\kgy}{kg$\cdot$y}

\newcommand{\kamland}{\mbox{{\sc k}am{\sc land-z}en}}

\begin{document}

\widetext

\title{A Combined Limit on the Neutrino Mass from Neutrinoless Double-$\beta$ Decay and Constraints
on Sterile Majorana Neutrinos}
\author{Pawel Guzowski}
\author{Luke Barnes}
\author{Justin Evans}
\author{Georgia Karagiorgi}
\author{Nathan McCabe}
\author{Stefan S\"oldner-Rembold}
\affiliation{The University of Manchester, Manchester M13 9PL, United Kingdom}
\date{\today}

\begin{abstract}
We present a framework to combine data from the latest neutrinoless double-$\beta$ decay experiments for multiple isotopes and derive a limit
on the effective neutrino mass $m_{\beta\beta}$ using the experimental energy distributions.
The combined
limits on \mbb~range between $130-310$~meV, where the spread 
is due to different model calculations of nuclear matrix elements (NMEs).
The statistical consistency ($p$ values)
between this result and the signal observation claimed by the Heidelberg-Moscow experiment
is derived. The limits on \mbb~are also evaluated in a $(3+1)$ sterile neutrino model, assuming all neutrinos are Majorana particles. 
\end{abstract}

\pacs{23.40.-s,14.60.Pq,14.60.St}
\maketitle

\section{Introduction}

The observation of neutrinoless double-$\beta$ $(0\nu\beta\beta)$ decays would demonstrate the Majorana 
nature of neutrinos~\cite{furry}, representing direct evidence for physics beyond the standard model. 
Neutrinoless double-$\beta$ decay is a second-order electroweak process where a nucleus
decays through the emission of two electrons, $(A,Z) \to (A,Z+2)  + 2 e^-$, thereby violating lepton number. The half-life of the isotope decaying through
$0\nu\beta\beta$ decay is given by 
\begin{equation}
  [T_{1/2}^{0\nu}]^{-1} = G^{0\nu} |M^{0\nu}|^2 \frac{m_{\beta\beta}^2}{m_e^2} \; , \label{eqn:halflife}
\end{equation}
assuming the exchange of a light Majorana neutrino (mass mechanism).
Here, $m_e$ is the electron mass. The phase space factors $G^{0\nu}$ and nuclear matrix elements $M^{0\nu}$
vary with isotope.
The decay rate is proportional to the square of the effective neutrino mass $m_{\beta\beta}$, which 
represents a coherent sum over the masses $m_i$ of the neutrino mass eigenstates, 
\begin{equation}
m_{\beta\beta}= \left | m_1 \left|U_{e1}\right|^2+m_2 \left|U_{e2}\right|^2 e^{i\alpha}+ m_3  \left|U_{e3}\right|^2 e^{i\beta} \right| \, ,
\label{eq-mbb}
 \end{equation}
weighted by the squares of the
corresponding elements  $U_{ei}$ of the Pontecorvo Maki Nakagawa Sakata (PMNS) matrix and the Majorana phases $\alpha$ and $\beta$.

The seesaw mechanism is the most common model for neutrino mass
generation~\cite{seesaw}. Here, heavy right-handed
neutrinos mix with the left-handed neutrinos and generate
light Majorana masses for the active neutrinos.
In addition to the exchange of these light Majorana neutrinos,
there are several other potential models~\cite{deppisch} that could contribute to $0\nu\beta\beta$ 
decays, e.g.,
left-right symmetric models, $R$-parity violating supersymmetry, or models with extra 
dimensions. Observing $0\nu\beta\beta$ decays
with different isotopes could help to disentangle these mechanisms~\cite{SN}.

The search for $0\nu\beta\beta$ decays is 
pursued in a range of experiments that use different isotopes. 
Experiments where detector material and $0\nu\beta\beta$ isotope are identical
are constrained in the choice of isotope by the detector technology. 
This approach is employed by collaborations
such as {\sc gerda}~\cite{gerda}, which uses a high-purity \iso{Ge}{76} detector, {\sc exo-200}~\cite{exo} and 
\kamland~\cite{kamland}, which contain enriched \iso{Xe}{136}, and {\sc cuoricino}~\cite{cuoricino} and {\sc cuore-0}~\cite{cuore}, bolometers made of TeO$_2$ crystals containing \iso{Te}{130}.
Alternatively, detector and isotope can be separated, which allows experiments such as {\sc nemo-3}~\cite{nemo3} to study a variety of isotopes, e.g., \iso{Mo}{100} and \iso{Se}{82}.

It is therefore important to develop techniques
to combine the data taken by different experiments with a range of isotopes. This allows us
to make quantitative comparisons between experiments, to study the 
consistency of their results, and to obtain combined limits on $m_{\beta\beta}$. 
A previous comparison of different experimental results using the published limits on $T_{1/2}^{0\nu}$
has been performed in Ref.~\cite{addendum}. In this analysis, we simultaneously fit the experimental energy distributions 
to obtain a combined limit.

Such a direct combination of the results for multiple isotopes requires  a
specific NME calculation to relate the half-lives of the various isotopes before deriving a limit on $m_{\beta\beta}$.
We chose a set of commonly used NME models to perform the combinations, but the procedure is generally applicable to include any NME model.
The combination of the experimental energy distributions is performed for each particular NME model.
In principle, different models
could give a better estimate of the NMEs for different isotopes, but this case is not considered here, since 
it would require determining
systematic correlations between the different models. The final limits on \mbb~are 
given as a range covering the full set of NME models, following the procedure normally employed by individual experiments.
We take into account theoretical uncertainties on the calculations where they are available.

In this Article, 
we present a method based on the published energy distributions
from all recent experiments and use it to derive a first combined limit on $m_{\beta\beta}$ based
on multiple isotopes. 
We first demonstrate that we can reproduce the published half-life limits of each individual
experiment, before performing the combination using different NME calculations. We 
study the consistency of the combined  $m_{\beta\beta}$ limit with the positive claim of the Heidelberg-Moscow
experiment~\cite{HM}. We also interpret the combined limit on $m_{\beta\beta}$ as a constraint on
a $(3+1)$ model with three active and one sterile Majorana neutrinos.

\section{Method}
\label{sec:method}

Limits are calculated using the  {\it CL}$_{s}$ method~\cite{Junk:1999kv,Read:2002hq,Fisher:2006zz}, which applies a modified-frequentist approach using a Poisson log-likelihood ratio (LLR) test statistic. The value of  {\it CL}$_{s}$ is defined as the $p$ value of the data under the hypothesis that we
observe both signal and background, {\it CL}$_{sb}$, divided by the $p$ value for the background-only hypothesis ({\it CL}$_b$). Systematic uncertainties are marginalized through Gaussian constraints on the priors, with the best fits of these parameters determined by maximizing the likelihood with respect to the data in both the signal-plus-background and background-only hypotheses. A limit at the $90\%$~Confidence Level (CL) is obtained from the signal strength that produces a value of  {\it CL}$_{s}=(100-90)\%$. The expected limit corresponds to the median of an ensemble of pseudo-experiments generated from the background-only probability distributions, and the $\pm 1$ standard deviation ranges are derived from the ensemble. When combining multiple experiments, the limit setting process uses a sum of the experiments' individual LLRs, assuming no correlations.

For a single experiment, the signal strength is inversely proportional to the half-life $T_{1/2}^{0\nu}$.
When combining multiple experiments, 
the signal strength is calculated for a common $m_{\beta\beta}$, which is related to $T_{1/2}^{0\nu}$ using Eq.~\ref{eqn:halflife} and 
for a specific NME calculation. 

\section{Experimental Inputs}

\begin{figure*}
    \centering
    \includegraphics[width=0.465\textwidth]{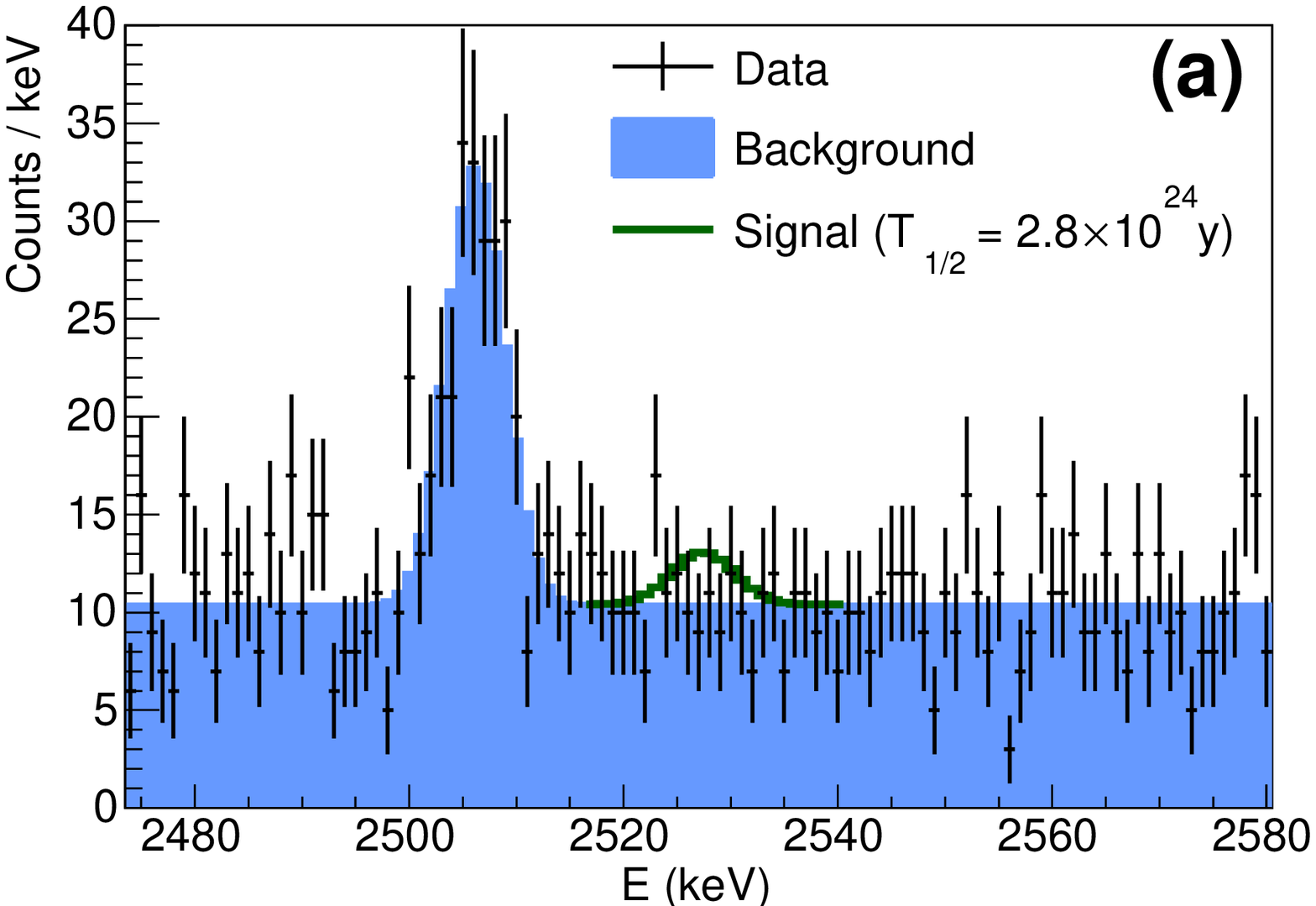} 
    \includegraphics[width=0.465\textwidth]{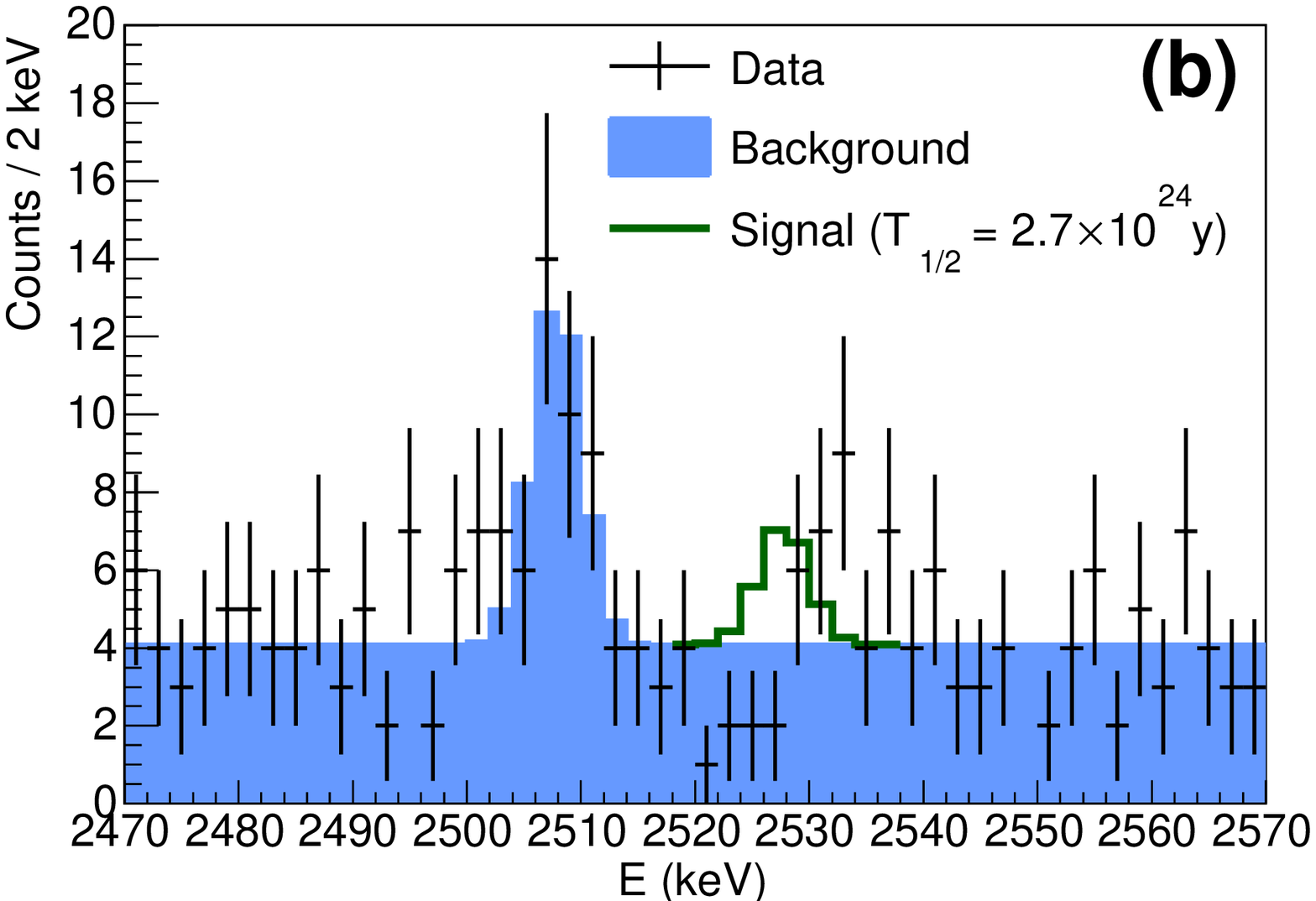}    
     \includegraphics[width=0.465\textwidth]{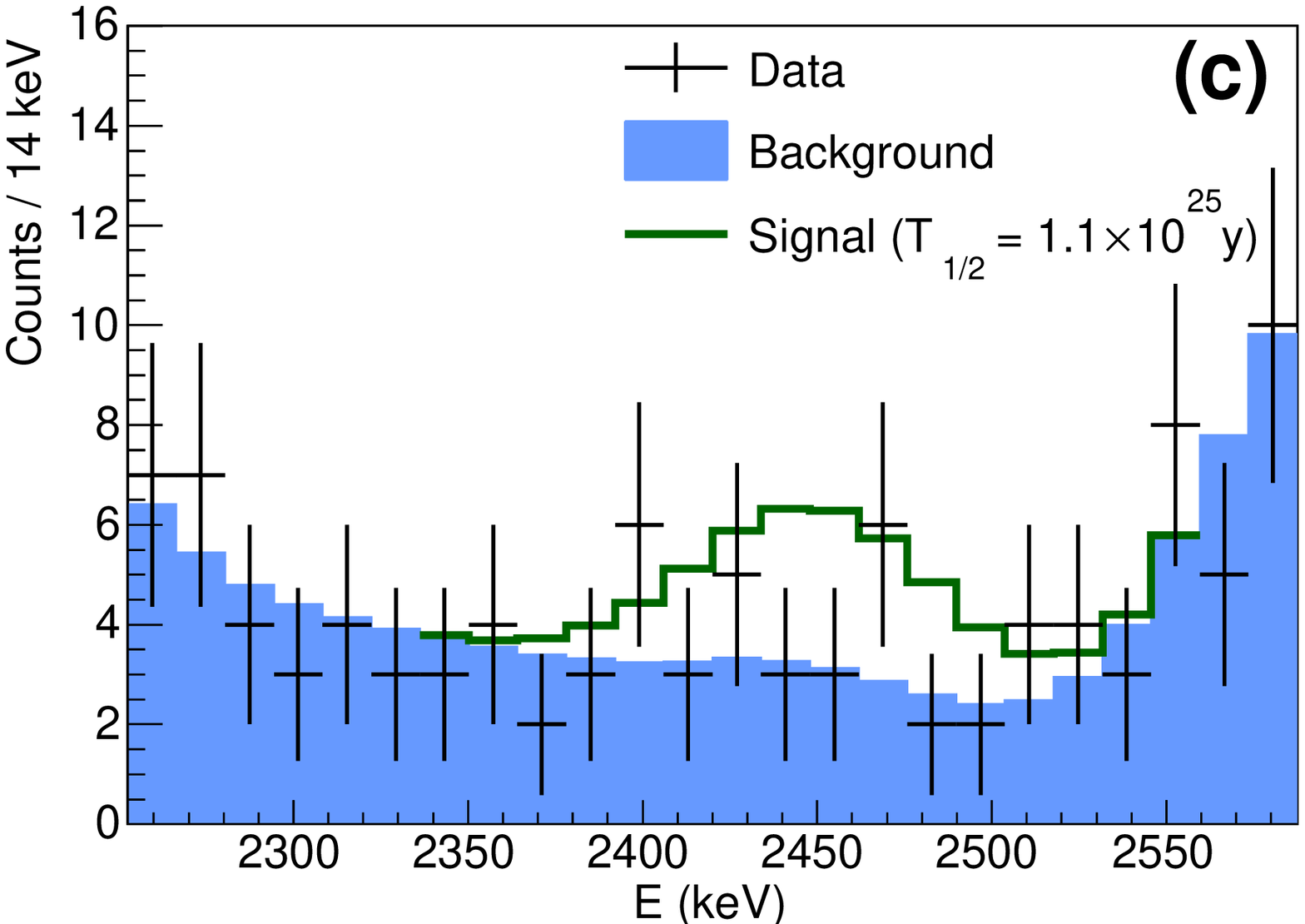}
    \includegraphics[width=0.465\textwidth]{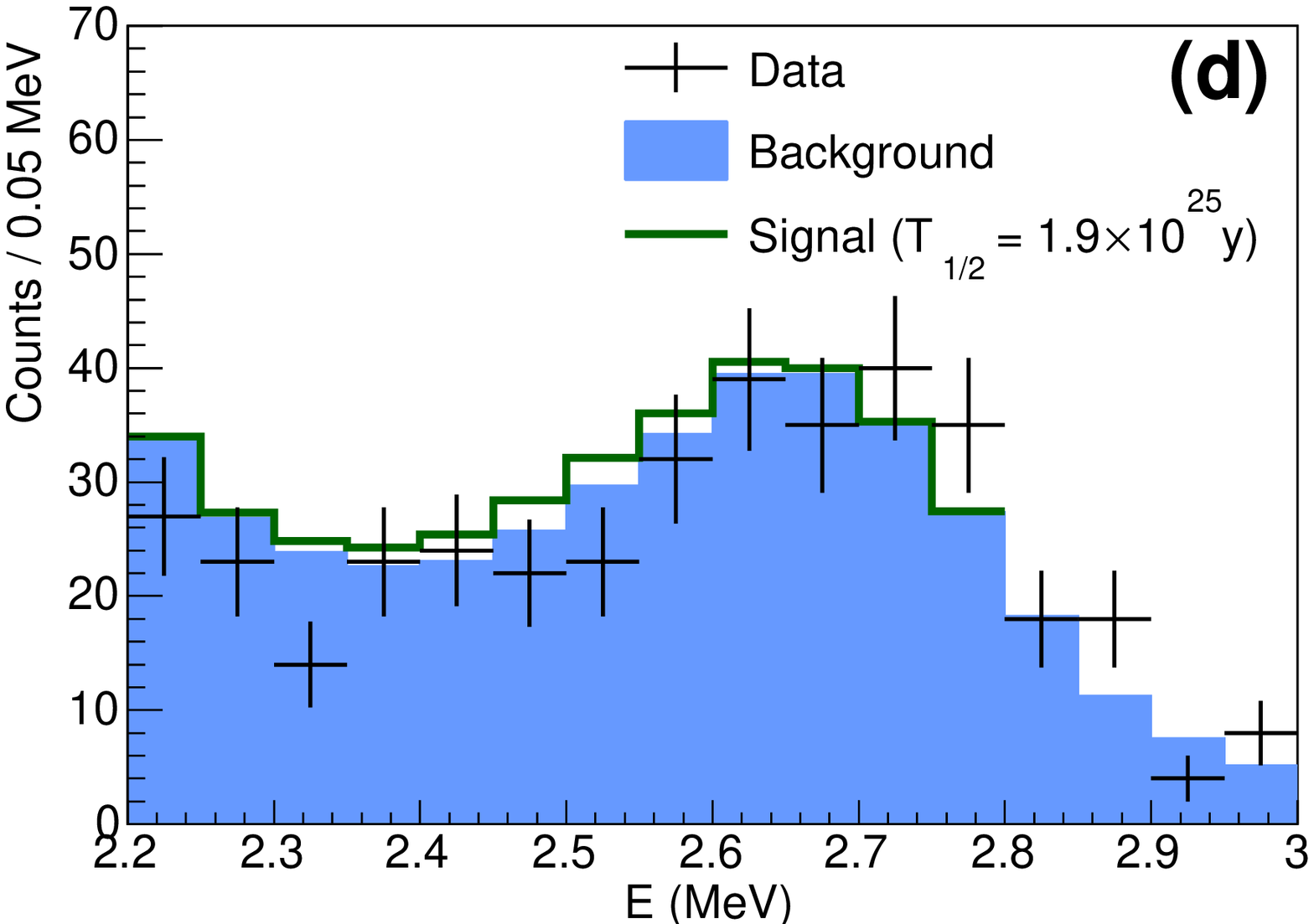}  
    \includegraphics[width=0.465\textwidth]{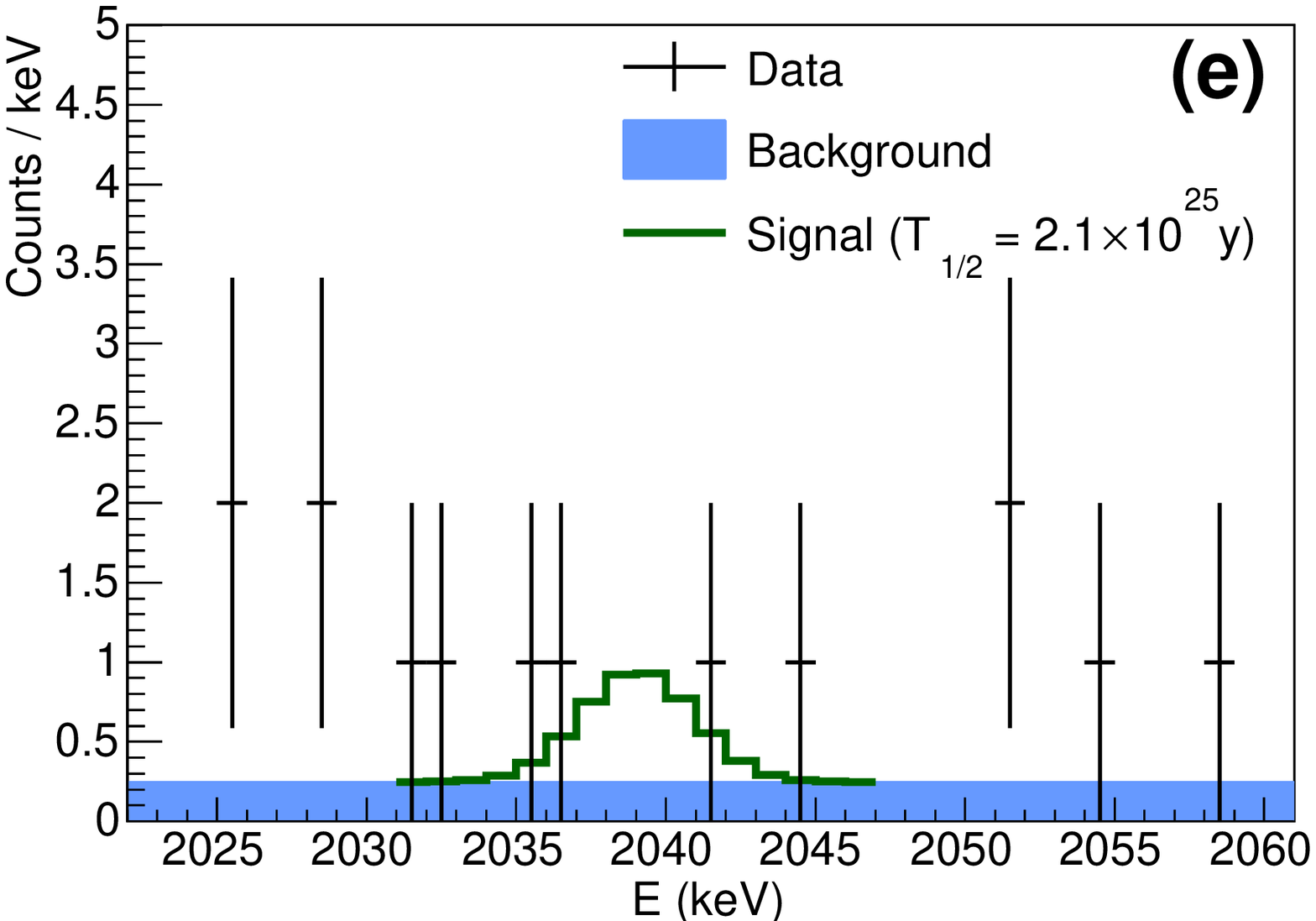} 
    \includegraphics[width=0.465\textwidth]{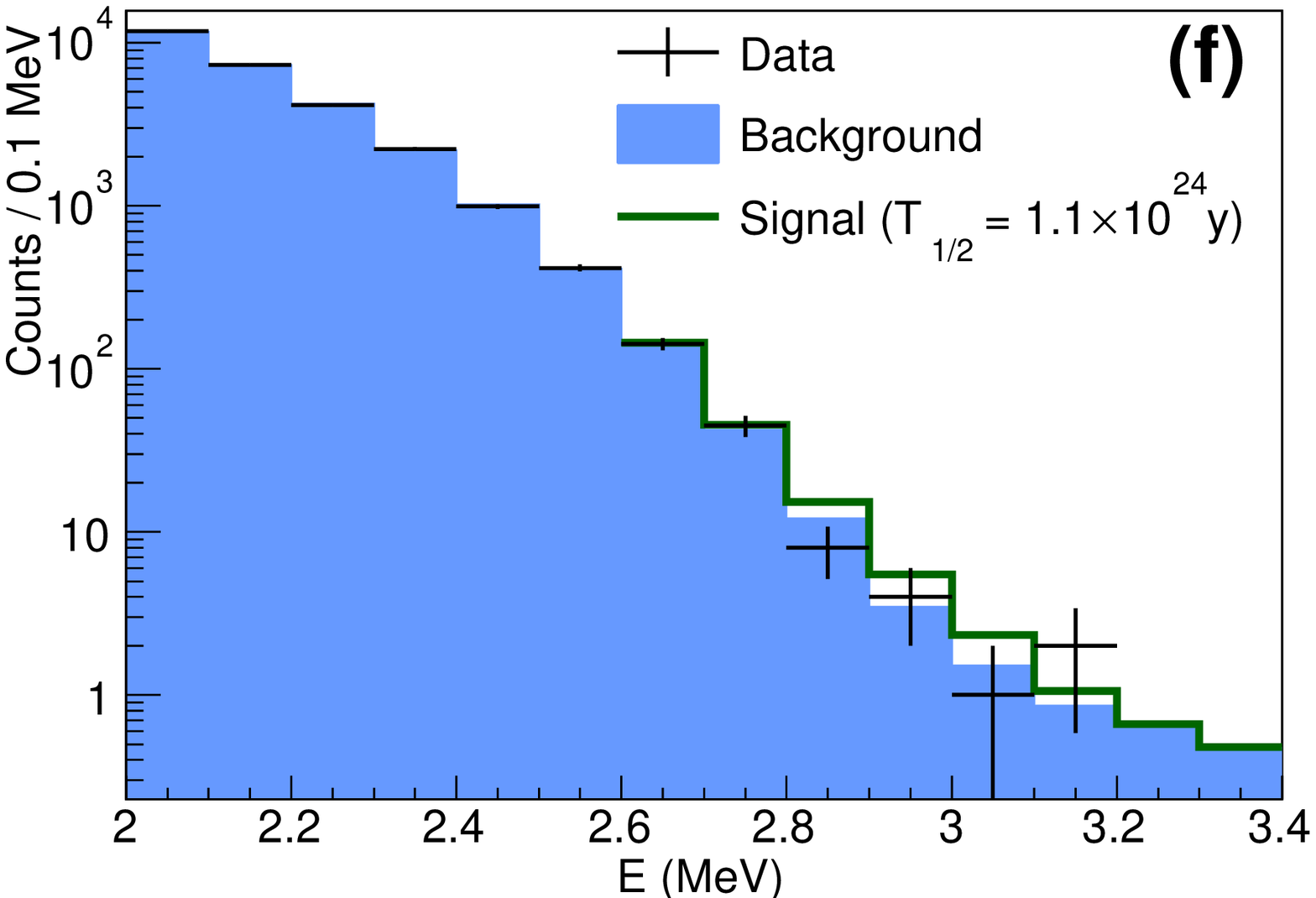} 
  \caption{Energy distributions for signal and background used as input for the combination with their statistical uncertainties:
  (a) {\sc cuoricino}, (b) {\sc cuore-0}, (c)  {\sc exo-200}, (d)  \kamland,  (e) {\sc gerda}, and (f) {\sc nemo-3} data.
  The signal distributions are normalized to represent the $90\%$ CL observed limits.}
  \label{fig:inputs}
\end{figure*}

The most stringent published limits on neutrinoless double-$\beta$ decay are currently provided by the {\sc cuoricino}, {\sc cuore},
{\sc exo-200}, 
\kamland, {\sc gerda}, and {\sc nemo-3} experiments. We use their most recent published energy distributions, together with the statistical and
systematic uncertainties
on signal and backgrounds, and the correlations of the systematic uncertainties 
as quoted by the experiments. The input distributions are shown in Fig.~\ref{fig:inputs} with
their statistical uncertainties. In total, we combine 250 data points. The experimental systematic uncertainties are assumed to be 
uncorrelated between experiments, because the data are taken using completely different experimental approaches.

{\sc cuoricino} was a bolometric detector comprising 62 TeO$_2$ crystals, which were operated between 2003 and 2008, for a total exposure of $19.75$~\kgy~\cite{cuoricino}. 
Using this exposure and the average signal efficiency of $82\%$, we obtain
a total expected number of $18$ signal decays for a half-life of $2.8 \times 10^{24}$~years. 
The energy distribution for the signal published in Fig.~9 of Ref.~\cite{cuoricino} normalized to this decay rate is used as the input distribution
for the limit setting. We use an uncertainty for the signal normalization of $1.5\%$ and an energy scale uncertainty of 
$0.8$~keV. A normalization uncertainty of $4.2\%$ is applied for the constant background, and the \iso{Co}{60} background, which is centred at an energy
of $2505$~keV, has a normalization uncertainty of $7.7\%$. These 
 background uncertainties are the exposure-weighted statistical uncertainties taken from Table 4 of Ref.~\cite{cuoricino}.
 
The {\sc cuore} experiment  has also published results from a first phase, {\sc cuore-0}, which utilizes a single
tower with $52$ TeO$_2$ crystals. The results correspond to an exposure of
$9.8$~\kgy~for the isotope \iso{Te}{130}. We use the data published in Fig.~3 of Ref.~\cite{cuore} and
the reported signal efficiency of $(81.3 \pm 0.6)\%$.
The signal shape is taken 
from Ref.~\cite{cuore2}, parametrized by a Gaussian function with width 
of $\approx 5$~keV. The statistical uncertainty on the background normalization is $3.45\%$, whereas
the effect of other systematic uncertainties is negligible. 

The {\sc exo-200} detector contains a liquid xenon time projection chamber that has been in operation since 2011~\cite{exo}. A total of 200 kg of xenon is used, with \iso{Xe}{136} enriched to $81\%$. We use the energy distribution of
the single-site (SS) decays published by the {\sc exo-200} Collaboration in Fig.~4(a) of 
Ref.~\cite{exo} for an exposure of 100 \kgy, since the sensitivity of SS topologies to $0\nu\beta\beta$ decays is significantly higher than for multi-site (MS) topologies.
We apply a signal normalization uncertainty of 8.6\% with an energy scale uncertainty of 0.2\%. The background normalization uncertainty is 10.9\%. We also apply an overall normalization uncertainty of 9.6\%, correlated between signal and background distributions, which is derived from the ratio
of SS to MS topologies. 

The \kamland~detector is filled with 13 tons of liquid scintillator, which is loaded  with enriched \iso{Xe}{136}~\cite{kamland}. We use Fig.~1(a) of Ref.~\cite{kamland}, published after a total exposure $89.5$~kg$\cdot$y with a background normalization uncertainty of $11.21\%$
and a signal normalization uncertainty of $3.99\%$.

The {\sc gerda} Collaboration uses high-purity germanium calorimeters enriched in \iso{Ge}{76}~\cite{gerda}. 
We use the dataset obtained between 2011 and 2013~\cite{gerda}, with an exposure of 21.6 \kgy. The energy distribution is taken from 
Fig.~1 of Ref.~\cite{gerda} after pulse shape discrimination.
We apply a 9\% uncertainty on the signal normalization and a 20\% uncertainty on the constant background normalization.

The {\sc nemo-3} Collaboration took data from 2003 to 2011 with a tracker-calorimeter detector measuring seven $0\nu\beta\beta$ candidate 
isotopes~\cite{nemo3}. 
Since  the \iso{Mo}{100} measurements are the most sensitive due to the larger isotope mass, we focus on this isotope.
The input distribution is the data shown in Fig.~2 of Ref.~\cite{nemo3}, which corresponds to an exposure of $34.7$ \kgy. The signal normalization uncertainty is $7\%$,
and the normalization uncertainty are $0.7\%$ for the two-neutrino background and $10\%$ for the other backgrounds.

\begin{table}[htbp]
\centering
\begin{tabular}{@{} lccccc @{}}
\hline\hline
 &  \multicolumn{4}{c}{Limit on $T^{0\nu}_{1/2}~(10^{24}$ y)} & \\
\cline{2-5}
 Experiment        & Publ. & Obs. & Exp. & $\pm 1 \sigma$ range & $1-${\sl CL}$_b$ \\
\hline
   \iso{Te}{130}:   & & & \\ 
 {\sc cuoricino}  &  2.8~\protect\cite{cuoricino}   & 2.8      & 2.9      & 2.0 -- 4.2      & 0.474    \\
  {\sc cuore-0}   &  2.7~\protect\cite{cuore}   & 3.0      & 3.0       & 2.1 -- 4.3      & 0.520    \\
 Combined & 4.0~\protect\cite{cuore} & 4.4       & 4.3       & 2.9  -- 6.2 & 0.513   \\
 \hline
    \iso{Xe}{136}:  & & & \\
 {\sc exo-200}         & 11~\protect\cite{exo}     & 13       & 21       & 14  -- 30       & 0.131    \\
 \kamland  &  19~\protect\cite{kamland}    & 17       & 11       & \phantom{0}7   -- 15       & 0.918    \\
 Combined & --- & 21       & 24       & 16  -- 34       & 0.360    \\
 \hline
   \iso{Ge}{76}:  & & & \\
  {\sc gerda}        & 21~\protect\cite{gerda}    & 20       & 21       & 14  -- 29       & 0.450    \\
  \hline
     \iso{Mo}{100}:  & & & \\
 {\sc nemo-3}     & 1.1~\protect\cite{nemo3}     & 1.1      & 0.9      & 0.6 -- 1.4      & 0.634    \\
\hline\hline
\end{tabular}
\caption{The published limits 
on $T^{0\nu}_{1/2}$ for each experiment are compared to the calculated observed and 
expected limits. The $\pm1\sigma$ range around the expected limit and the $1-${\sl CL}$_b$ value of the data are also shown. }
\label{tab:sum}
\end{table}

We expect small differences between our method and the published results, since we use a different limit setting procedure 
and only a single 
input distribution for each experiment. 
As a first step designed to validate our method, we determine the lower limits on the half-lives at the $90\%$ CL and compare them to the values
published by the collaborations (see Table~\ref{tab:sum}).
The results for \iso{Ge}{76} and \iso{Mo}{100} are in very good agreement
with the published values. 
The half-life limit for the {\sc exo-200} data are $15\%-20\%$ higher with our limit setting method, whereas
the result for \kamland~is about $10\%$ lower. We observe the same effect for a previously published {\sc exo-200} 
result~\cite{auger}, which has also been used by the \kamland~Collaboration to set a combined limit for \iso{Xe}{136}~\cite{kamland}. We combine the same two \iso{Xe}{136} data sets and obtain
a limit of $T_{1/2}^{0\nu}=34 \times 10^{24}$~y, in perfect agreement with the published \kamland~combination.

We also obtain perfect agreement for the {\sc cuoricino} result using \iso{Te}{130}, while the limit on $T_{1/2}^{0\nu}$ 
obtained by the {\sc cuore-0} Collaboration is $10\%$ lower than the limit we derive.  Since {\sc cuoricino} and {\sc cuore-0} are two phases of the
same experiment, we only use the combined {\sc cuore} limit to perform our remaining analysis.

The observed limits are also compared to the median expected limits and the corresponding $\pm 1\sigma$ range around
the median expected limits. 
All observed limits lie within the $\pm 1\sigma$ band, except for the \kamland~result, where the observed limit of
$T^{0\nu}_{1/2}= 17 \times 10^{24}$~y is about $1.5\sigma$ better than the expected limit,
corresponding to $1-${\sl CL}$_b$ of $0.918$. This is caused by the deficit of data compared to the background expectation in
the signal region.

\section{Nuclear Matrix Elements}

\begin{table*}[htbp]
   \centering
   \begin{tabular}{@{} c|c|cccccccccccccc @{}} 
    \hline\hline
     Isotope & Phase Space &  \multicolumn{13}{c}{Nuclear Matrix Element Models} \\
       &Factor $G^{0\nu}$& GCM &  IBM-2& 
       NSM & \multicolumn{4}{c}{QRPA~\protect\cite{QRPA}} & pnQRPA~\protect\cite{pnQRPA}&   \multicolumn{6}{c}{(R)QRPA~\protect\cite{RQRPA}}  \\
      &  ($10^{-14}$y$^{-1}$)~\cite{Kotila:2012zza}  & \protect\cite{Rodriguez:2010mn}  & \protect\cite{ibm-2}  & \protect\cite{Menendez:2008jp}  & A-old & A-new & B-old & B-new  &   & NME  & Rel.~Unc. &
    \multicolumn{4}{c}{Correlation Matrix}     \\
              \hline
      \iso{Ge}{76}  & 0.615  & 4.60 & 4.68 & 2.30 & 5.812 & 5.157 & 6.228 & 5.571 & 5.26 & 4.315 & 0.191     & 1     &       &       &   \\ 
      \iso{Mo}{100} & 4.142  & 5.08 & 4.22 & ---  & 5.696 & 5.402 & 6.148 & 5.850  &  3.90 & 3.184 & 0.254    & 0.973 & 1     &       &   \\
      \iso{Te}{130}  & 3.699   & 5.13 & 3.70 & 2.12 & 4.306 & 3.888 & 4.810 & 4.373  & 4.00 & 3.148 & 0.247    & 0.899 & 0.862 & 1     &   \\ 
      \iso{Xe}{136}   & 3.793 & 4.20 & 3.05 & 1.76 & 2.437 & 2.177 & 2.735 & 2.460 & 2.91 & 1.795 & 0.293   & 0.805 & 0.747 & 0.916 & 1 \\
     \hline
   \end{tabular}
   \caption{Phase space factors for $g_A=1.27$ for the four isotopes,
     values of the nuclear matrix elements for the GCM, IBM-2, NSM, QRPA, pnQRPA,
     and for the (R)QRPA NME calculation, together with 
     the relative uncertainties on the (R)QRPA NMEs and their correlation matrix. The relative uncertainties
     quoted for the IBM-2 calculation are $0.16$ for each isotope.
  }
   \label{tab:params}
\end{table*}

We derive limits on the effective neutrino mass for different NME calculations, which are summarized in 
Table~\ref{tab:params}.
The five different models used in the combination are the
Generating Coordinate Method (GCM)~\cite{Rodriguez:2010mn}, 
the Interacting Boson Model (IBM-2)~\cite{ibm-2}, the
Interacting Shell Model (NSM)~\cite{Menendez:2008jp}, and three models using the 
Quasiparticle Random-Phase Approximation: (i) the QRPA model
using four different sub-calculations based on the Argonne V18 (A) or charge-dependent Bonn (B) nucleon-nucleon potentials with an older (\emph{old}) or newer (\emph{new}) version of the parametrization of the particle-particle interactions~\cite{QRPA}, (ii) the 
proton-neutron QRPA model (pnQRPA)~\cite{pnQRPA},
and (iii) a QRPA model that uses its renormalized version for evaluating uncertainties, labelled (R)QRPA~\cite{RQRPA}.

The model calculations predict values for the NMEs which can differ by up to a factor of
$\approx 2$ for a particular isotope. The NSM model predicts the smallest NMEs for all isotopes, apart from \iso{Mo}{100} which is not
evaluated in this calculation, whereas the models differ in 
their predictions of the isotopes with the largest NMEs.

The (R)QRPA and the IBM-2 calculations also provide uncertainties on the NMEs. The IBM-2 model uncertainties on the NMEs
are $16\%$, which we have assumed to be fully correlated across isotopes.
The (R)QRPA model includes a correlation matrix between isotopes, which we take into account in the limit setting.  The uncertainties are evaluated 
in Ref.~\cite{RQRPA}
by using (i) two values of the  weak axial-vector coupling parameter, $g_A=1$ and $1.25$, 
(ii) two different approaches to short-range correlations, 
(iii) two many-body models, QPRA and RQPRA,  and (iv) three different sets of single-particle states. 
The uncertainties also include the experimental uncertainties on the particle-particle coupling constant 
$g_{pp}$ extracted from $2\nu\beta\beta$ data. The resulting 24 combinations are used to extract correlation coefficients
from the error ellipses which cover the full range of possible outcomes. For our experimental combination, we reinterpret these model uncertainties
 as Gaussian uncertainties by reducing them by a factor of $0.68$.
We quote the reduced uncertainties in Table~\ref{tab:params}.

The phase space factors $G^{0\nu}$ are taken from a recent 
calculation~\cite{Kotila:2012zza}. We 
set the weak axial-vector coupling parameter to $g_A=1.27$. The uncertainties on the phase space factors
originate from the uncertainties on the $Q$ values and the nuclear radius.  
The uncertainty on the nuclear radius
dominates and leads to an uncertainty on the phase space factor $G^{0\nu}$ of $\approx 7\%$.  

\section{Results}

Only in the case of \iso{Xe}{136}, two independent experiments, \kamland~and {\sc exo-200}, 
measured the same isotope.
Their data can be combined directly without using NMEs to derive a limit on $T^{0\nu}_{1/2}$. The combination yields
a limit of $T^{0\nu}_{1/2}>2.1 \times 10^{25}$~y, which is dominated by the {\sc exo-200} measurement due to the
higher exposure compared to \kamland. This observed value is less stringent than the previously 
published combined value of $T^{0\nu}_{1/2}>3.4 \times 10^{25}$~y~\cite{kamland}
based on a smaller {\sc exo-200} data set~\cite{auger}, which displays a downward fluctuation of the data relative to the background
expectation.
The combined sensitivity, given by the median expected limit, has improved
from $T^{0\nu}_{1/2}=1.6\times10^{25}$ y to $T^{0\nu}_{1/2}=2.4\times10^{25}$~y due to the increased exposure.

\begin{figure}[htbp]
   \centering
   \includegraphics[width=0.48\textwidth]{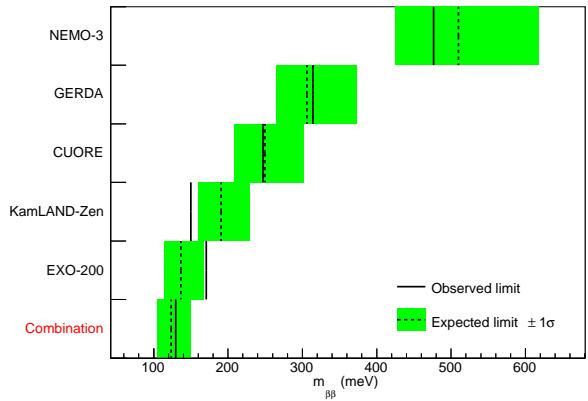} 
   \caption{Individual experiment effective mass limits, and the combined limit, using the
   GCM model.}
   \label{fig:gcm}
\end{figure}

To combine results for multiple isotopes, we include the effects of NMEs. 
We show an example in Fig.~\ref{fig:gcm} for the GCM calculation.
The results for $m_{\beta\beta}$ are shown for each individual experiment, ordered by their sensitivities
using this model.
The combined observed limit is $m_{\beta\beta}<130$~meV for an expected limit of 
$m_{\beta\beta}<130$~meV, which 
improves upon the best individual limit obtained by \kamland~of $m_{\beta\beta}<150$~meV 
and the best individual expected limit by {\sc exo-200} of $m_{\beta\beta}<140$~meV.
All limits are given at the $90\%$ CL.
Since the mass limit depends on the fourth root of the exposure, these improvements corresponds to an increase in exposure by a factor of $1.4$--$1.6$. 

Using the GCM model, the {\sc cuore} and {\sc gerda} data have sensitivities of $m_{\beta\beta}=250$~meV and
$310$ meV, respectively, while the two experiments using \iso{Xe}{136} have the best sensitivities in the range
$m_{\beta\beta}=140$--$190$~meV.
\begin{figure}[htbp]
   \centering
   \includegraphics[width=0.48\textwidth]{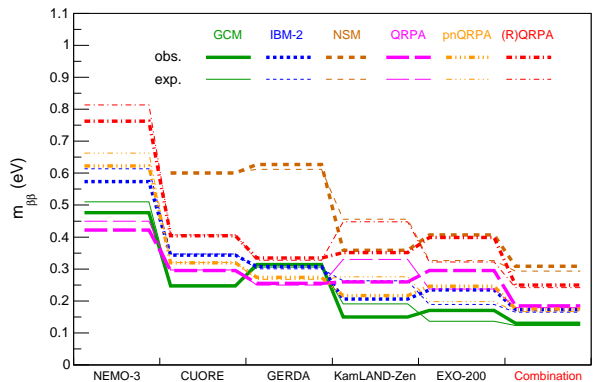} 
   \caption{Observed and expected limit  on \mbb~for the different experiments and NME models without NME model uncertainties.}
   \label{fig:nmeall}
\end{figure}
Figure~\ref{fig:nmeall} shows the observed and expected limits for each NME model and experiment separately. 
For this comparison, NME uncertainties are not taken into account, since only
some of the calculations provide them. The ordering
of the experimental sensitivities changes with the NME model used to derive the \mbb~limit, further
emphasising the need for using a range of isotopes to search for $0\nu\beta\beta$ decays.

The coefficients of the correlation matrix provided by the (R)QRPA calculations for the four isotopes
considered here range from $0.747$ to $0.973$ (see Table~\ref{tab:params}). 
To study the effect of these correlations, we derive the limits on $m_{\beta\beta}$  assuming no NME uncertainties,
the full correlation matrix, and a single correlation coefficient
for all isotope combinations ranging from $-0.3$ to $1$. 
\begin{figure}[htbp]
   \centering
   \includegraphics[width=0.48\textwidth]{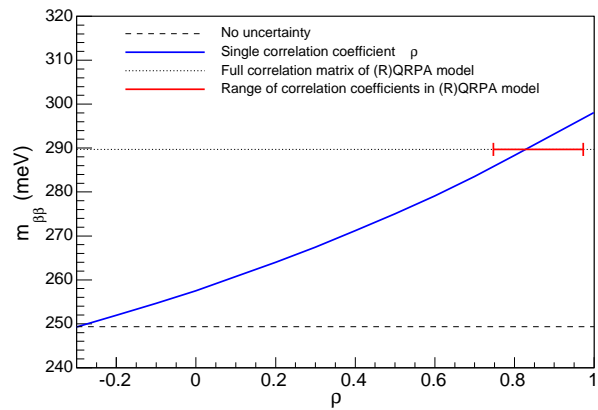} 
   \caption{Limit on $m_{\beta\beta}$ as a function of a single correlation coefficient $\rho$ in the (R)QRPA model.
    The lines at  $m_{\beta\beta}=250$~meV and $290$~meV represent the limits without uncertainties and 
     with the full correlation matrix of the (R)QRPA model, respectively. The red line shows the range of 
     correlation coefficients.}
   \label{fig:correl}
\end{figure}
The results are shown in Fig.~\ref{fig:correl}. 
Using no NME uncertainties yields a limit of $m_{\beta\beta}<250$~meV, which increases
to $m_{\beta\beta}<290$~meV once the model uncertainties and the full correlation matrix
are taken into account, and to $m_{\beta\beta}<300$~meV assuming $100\%$ correlation between
the NME uncertainties for the different isotopes. 
The variation in limits of $\approx 40-50$~meV is comparable to the range observed for the different model
assumptions with the QPRA calculation, which are expected to represent similar model
uncertainties.

\begin{table}[htbp]
   \centering
   \begin{tabular}{lccccc} 
      \hline\hline
        NME         & $m^{\rm obs}_{\beta\beta}$  & $m^{\rm exp}_{\beta\beta}$  &    \multicolumn{2}{c}{Improvement} & $p$ value\\
           &  (meV) & (meV) &  Limit  & Sensitivity &   (HM)\\
      \hline
      GCM & 130 & 120 & 14\% (K) & 10\% (E) & 0.001\\
        IBM-2: & & & & \\
      no unc. & 170 & 170 & 16\% (K) & 12\% (E) & 0.010\\
      with unc. & 190 &  180 & 16\% (K) & 13\% (E) & 0.021\\
      NSM & 310 & 290 & 14\% (K) & 10\% (E) & 0.003\\
         QRPA: & & & & \\
      A-new & 200 & 200 & 23\% (G) & 25\% (E) & 0.095\\
      A-old & 180 & 180 & 26\% (G) & 25\% (E) & 0.100\\
      B-new & 180 & 180 & 28\% (K)  & 24\% (E)  & 0.073\\
      B-old & 170 & 160 & 28\% (K)  & 23\% (E) & 0.077\\
        pnQRPA & 170 & 170 & 19\% (K) & 16\% (E) & 0.029\\
       (R)QRPA: & & & & \\
      no unc. & 250 & 240 & 25\% (G) & 25\% (E) & 0.109\\
      with unc. & 290 & 290 & 23\% (G) & 21\% (G) & 0.311\\
      \hline\hline
   \end{tabular}
   \caption{For each NME calculation, the combined observed and expected limits on \mbb, the improvement in the limit and the 
   sensitivity relative to the best individual experiments for that NME model. The
   best experiments are {\sc gerda} (G), {\sc exo-200} (E), or \kamland~(K).
   The $p$ value of the limit with respect to the Heidelberg-Moscow (HM) positive claim~\cite{HM} are also shown. 
   All \mbb~limits are given to two significant digits.}
   \label{tab:combis}
\end{table}
In Table~\ref{tab:combis}, we show the combined limits on
$m_{\beta\beta}$ for each NME model. 
We provide the observed and expected limits for each combination and the improvements in observed limit and sensitivity 
over the result obtained by the best individual experiment. 
As shown in Fig.~\ref{fig:nmeall}, the most sensitive experiments change for the different models.

\begin{figure}[tbp]
   \centering
   \includegraphics[width=0.48\textwidth]{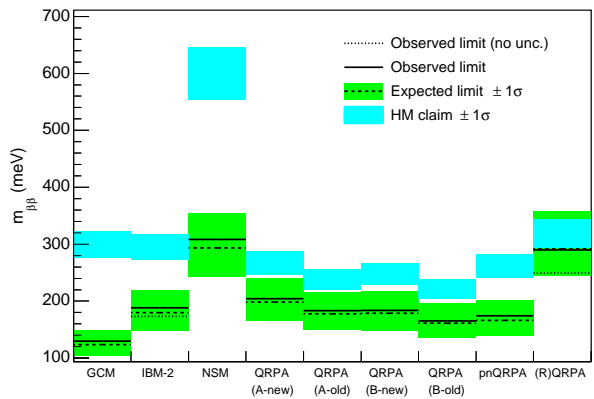} 
   \caption{Mass limits (observed and expected) for each NME calculation. 
   For the IBM-2 and (R)QRPA models, the limits are given both with and without the uncertainty on the NME calculation.
   The effective mass for the positive Heidelberg-Moscow (HM) claim~\protect\cite{HM} is also shown.}
   \label{fig:all}
\end{figure}
The combined
results are compared to the result obtained by the Heidelberg-Moscow (HM) experiment in Fig.~\ref{fig:all},
where the error bands represent the uncertainties of the HM half-life measurement of 
$T^{0\nu}_{1/2}=(2.23^{+0.44}_{-0.31}) \times 10^{25}$~y for \iso{Ge}{76}~\protect\cite{HM}.  
The $p$ values, testing
the consistency between the result obtained by the HM experiment using \iso{Ge}{76} and the combined upper
limits, are also given in Table~\ref{tab:combis}. 
They are calculated taking into account the experimental uncertainties.

The $p$ values differ significantly between different
NME models, and any conclusion on the level of consistency between the combination of
the latest experimental results and the result obtained with the 
the $T^{0\nu}_{1/2}$ reported by the HM group depends strongly on the chosen NME model.
The $p$ values range from $0.001$ (about $3$ standard deviations) for the GCM model
to $0.311$ for the (R)QRPA models. The $p$ value for the {\sc gerda} result
and this HM measurement is $0.63$, independent of NME model since the experiments both
use the isotope \iso{Ge}{76}.  

\begin{figure}[htbp]
   \centering
   \includegraphics[width=0.48\textwidth]{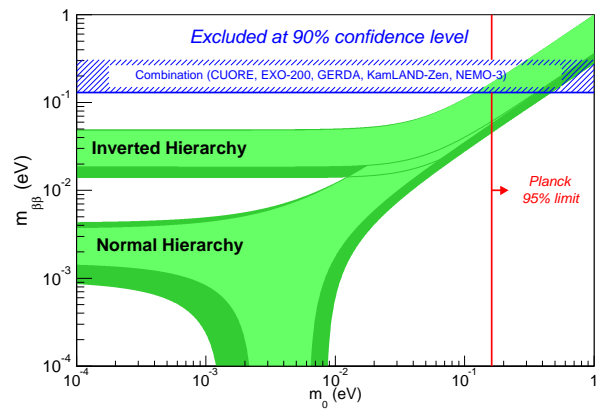} 
   \caption{Combined limits on the effective mass \mbb~as a function of the lightest neutrino mass $m_0$. The width of the horizontal band represents the range of limits obtained with the different NME models.   
   The neutrino oscillation parameters are taken from~Ref.~\protect\cite{Gonzalez-Garcia:2014bfa}, using the best fit values. The inner band for the NH and IH fits shows the full range associated with the unknown Majorana phases and the
   outer band the additional effect of the $3$-standard deviation experimental uncertainties on the oscillation parameters.
   The limit on $m_0$ derived from the Planck data is also shown.}
   \label{fig:hierarchies}
\end{figure}

The best published limit on $m_{\beta\beta}$ by \kamland~\cite{kamland} is obtained by combining their data with the earlier 
{\sc exo-200} result~\cite{auger}. Using a similar set of NMEs, the mass range they obtain is 
$m_{\beta\beta}=120$--$250$~meV, whereas our observed combined limit in the range $m_{\beta\beta}=130$--$310$~meV
is slightly higher because we use the latest {\sc exo-200} data set. 

In Fig.~\ref{fig:hierarchies}, we overlay the range of observed combined limits on the allowed effective mass \mbb~as a function of the lightest neutrino mass~$m_0$, with the bands representing current measurement uncertainties of the neutrino oscillation parameters assuming either the normal hierarchy (NH) with $m_1<m_2<m_3$ or the inverted hierarchy (IH) with $m_3<m_1<m_2$ for the ordering of the neutrino mass eigenstates
in the standard three-neutrino framework. We also show a limit on $m_0$, which is derived from the
limit on the sum of the neutrino masses obtained by the 
Planck Collaboration~\cite{Planck}. They set a limit of  
$\sum_i m_i<492$~meV at the $95\%$ CL  based on the temperature
and the polarization of the Cosmic Microwave Background.

We also interpret the combined limit on $m_{\beta\beta}$ as a constraint on a single sterile Majorana neutrino that mixes with the three active Majorana states. In this model, the expression for the effective neutrino mass $m_{\beta\beta}$ defined in Eq.~\ref{eq-mbb} is modified
to take into account the mixing with the fourth neutrino state:
\begin{eqnarray}
\lefteqn{m_{\beta\beta}  =  \left| m_1 \left|U_{e1}\right|^2+m_2 \left|U_{e2}\right|^2 e^{i\alpha}+ \right.} \nonumber \\
& &  \;\;\;\;\;\;\;\;\;\;  \left. + m_3  \left|U_{e3}\right|^2 e^{i\beta}+ m_4 \left|U_{e4}\right|^2 e^{i\gamma} \right| .
\end{eqnarray}
It includes an additional mass term $m_4$, a Majorana
phase $\gamma$, and the element $U_{e4}$ of the now extended, $4\times 4$ mixing matrix, which
is related to the sterile neutrino mixing angle $\theta_{14}$ through $\left|U_{e4}\right|^2 =\sin^2\theta_{14}$ 
in the parametrization used here. In this model, the unitarity constraint applies only for the $4\times 4$ mixing matrix. However, the central
values of the elements $U_{ei} (i=1,2,3)$ do not change significantly when the $3\times 3$ unitarity constraint is 
removed from the global fit~\cite{parke}. The  unitarity constraint on the $4\times 4$ mixing matrix will restrict 
$|U_{e4}|^2=\sin^2\theta_{14}$ to be $\lesssim 0.1$.  

Our ability to probe a certain range of $m_4\left|U_{e4}\right|^2$ depends on $m_0$ and the Majorana phases~(see, e.g., \cite{Li:2011ss,Girardi:2013zra}). 
We take the approach of translating the limits on $m_{\beta\beta}$ into corresponding limits on the combination $m_4\left|U_{e4}\right|^2$ under two extreme situations for a given $m_0$:
\begin{itemize}
\item The NME model (NSM) that predicts the highest (least stringent) limit on $m_{\beta\beta}$, together with the phases $\alpha=\beta=0$, and $\gamma=\pi$  that provide the smallest contribution to $m_{\beta\beta}$.
\item The NME model (GCM) that predicts the lowest (most stringent) limit on  $m_{\beta\beta}$, together with the phases $\alpha=\beta=\gamma=0$ that provide the largest contribution to $m_{\beta\beta}$. 
\end{itemize}
 This first parameter set allows for large contributions from $m_4\left|U_{e4}\right|^2$ and thus provides the least constraining limit on $m_4\left|U_{e4}\right|^2$, whereas the second set allows for only small contributions from $m_4\left|U_{e4}\right|^2$ and thus yields the most constraining limit on $m_4\left|U_{e4}\right|^2$. 
 
\begin{figure}[htbp]
   \centering
   \includegraphics[width=0.48\textwidth]{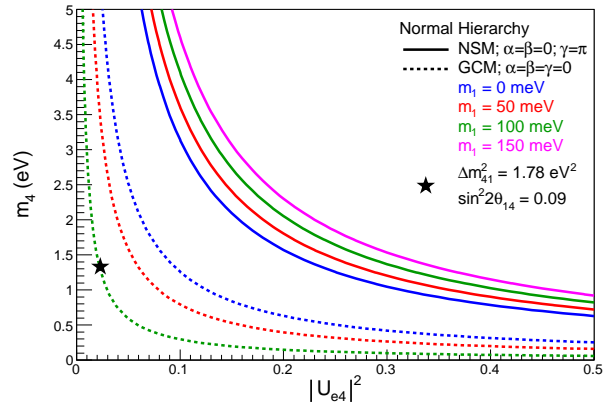} 
   \caption{Limits on $m_{\beta\beta}$ for different NME models translated into a constraint on the sterile neutrino
     in the $(m_4,\left|U_{e4}\right|^2)$ plane in a $(3+1)$ model.  
    Different values of the Majorana phases $\alpha,\beta,\gamma$ and 
  of the lightest active neutrino mass $m_1$ in the NH  are shown. The complete region
  is excluded for $m_1=150$~meV with the GCM model and vanishing Majorana phases.  }
   \label{fig:sterile_nh}
\end{figure}
\begin{figure}[htbp]
   \centering
   \includegraphics[width=0.48\textwidth]{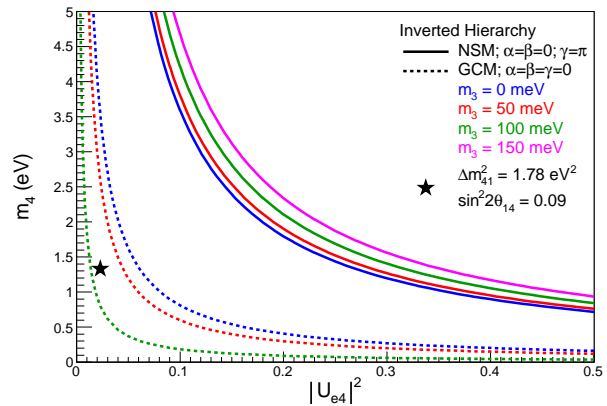} 
   \caption{Limits on $m_{\beta\beta}$ for different NME models translated into a constraint on the sterile neutrino
     in the $(m_4,\left|U_{e4}\right|^2)$ plane in a $(3+1)$ sterile neutrino model.  
    Different values of the Majorana phases $\alpha,\beta,\gamma$ and 
  of the lightest active neutrino mass $m_3$ in the IH  are shown.  The complete region
  is excluded for $m_3=150$~meV with the GCM model and vanishing Majorana phases.}   
   \label{fig:sterile_ih}
\end{figure}

In Figs.~\ref{fig:sterile_nh} and~\ref{fig:sterile_ih}, we translate the measured
combined limit on $m_{\beta\beta}$ for the two situations above  
into a limit in the ($m_4,\left|U_{e4}\right|^2$) plane for a set of different assumptions on $m_0$. To be consistent with cosmological bounds, 
we vary $m_0$ between $0$ and $150$~meV.  
For any of the $m_0$ values considered, 
all other NME model and Majorana phase combinations will produce
limit curves that lie in between these extremes. The values of $\Delta m^2_{ij}$ and
$U_{ei}$ $(i,j=1,2,3)$ are taken from a recent global fit~\cite{Gonzalez-Garcia:2014bfa}.
Including the three-standard deviation uncertainties on these fit values
has negligible effect on the results. 

The central value of $\sin^2 2\theta_{14}=0.09$ and $\Delta m^2_{41}= 1.78$~eV$^2$
from a global fit~\cite{bib:sterile_fit} of the oscillation parameters in a $(3+1)$ sterile neutrino model 
including reactor, gallium, solar neutrino, and {\sc lsnd/karmen} $\nu_e$ disappearance data, all of which are directly sensitive to $m_4$ and $\left|U_{e4}\right|$, is
also shown in Figs.~\ref{fig:sterile_nh} and~\ref{fig:sterile_ih}.  The limit from the $0\nu\beta\beta$ combined fit is currently close to the (3+1) global best fit value 
only in the more constraining scenarios calculated using the GCM model and with $m_0>100$~meV and favorable Majorana phases. 
Our results are consistent with previous analyses (see, e.g.,~\cite{Girardi:2013zra,Giunti:2012tn}) that investigate $m_{\beta\beta}$ regions allowed by global fits to oscillation experiments under a (3+1) hypothesis.
Assuming the sterile neutrino is a Majorana particle, 
neutrinoless double-$\beta$ decay can therefore independently constrain the existence of a sterile
neutrino under those extreme assumptions. For a more detailed discussion on constraints to (3+1) sterile neutrino models from neutrinoless double $\beta$ decay and projected implications for next-generation experiments, see Ref.~\cite{Giunti:2015}.

\section{Summary}

We have performed the first combination of the latest data sets from experiments searching for
neutrinoless double-$\beta$ decay  using
multiple isotopes.  Using the {\sl CL}$_s$ method, we set a limit on the
effective neutrino mass \mbb~in the range $130-310$~meV, depending on NME model.
Combining the data from multiple isotopes and experiments significantly increases the sensitivity over
using the single-best experiment only, corresponding to an increase by a factor of $\approx 1.5-2.4$
in exposure.
We compare these limits with the claimed observation of the Heidelberg-Moscow 
experiment and obtain $p$ values that differ significantly depending on the NME calculations chosen,
ranging from $0.001$ for the GCM model to $0.31$ for
the (R)QRPA model. Using the uncertainties and the full correlation matrix 
provided by the (R)QRPA model changes the limit on \mbb~by $40-50$~meV compared to using
no NME uncertainties. We also translate the combined limit on \mbb~into a constraint
on a light sterile Majorana neutrino in a $(3+1)$ model. This translated limit is NME and $m_0$ and Majorana phase dependent.

\section*{Acknowledgements}

We want to thank Fedor {\v S}imkovic and Sean Freeman for useful discussions about the nuclear matrix elements.

\end{document}